\documentclass[a4paper]{jpconf}
\usepackage{graphicx}
\usepackage{graphics}
\usepackage{iopams}
\usepackage{dsfont}
\usepackage{amsfonts}

\begin{document}
\newcommand{\half}{\mbox{$\textstyle \frac{1}{2}$}}
\newcommand{\quat}{\mbox{$\textstyle \frac{1}{4}$}}
\newcommand{\octa}{\mbox{$\textstyle \frac{1}{8}$}}
\newcommand{\rd}{{\rm d}}
\newcommand{\ri}{{\rm i}}
\newcommand{\re}{{\rm e}}

\title[Random Hamiltonian in thermal equilibrium]
{Random Hamiltonian in thermal equilibrium}

\author[D~C~Brody, D~C~P~Ellis, and D~D~Holm]
{Dorje~C~Brody, David~C~P~Ellis, and Darryl~D~Holm}

\address{Department of Mathematics, Imperial College London,
London SW7 2AZ, UK}

\begin{abstract}
A framework for the investigation of disordered quantum systems in
thermal equilibrium is proposed. The approach is based on a
dynamical model---which consists of a combination of a
double-bracket gradient flow and a uniform Brownian
fluctuation---that `equilibrates' the Hamiltonian into a canonical
distribution. The resulting equilibrium state is used to calculate
quenched and annealed averages of quantum observables.
\end{abstract}


\section{Introduction}
\label{sec:1}

In the conventional treatment of quantum statistical mechanics there
is a natural division between (a) the system under study, which is
treated quantum mechanically and whose states are subject to thermal
fluctuations, and (b) the Hamiltonian of the system, which is
treated essentially classically and is held fixed. For some quantum
systems, however, the Hamiltonian itself may fluctuate for one
reason or another. Questions that interest us in this connection, in
particular, are: ``How can a randomly fluctuating Hamiltonian
approach its equilibrium state?'' and ``What is the form of the
equilibrium distribution, and how do we calculate observable
expectation values in equilibrium?'' The former is a question of a
\textit{dynamical} nature, whereas the latter is a question of a
\textit{static} nature. The purpose of the present paper is to
propose an approach to address these questions. Specifically, we
shall derive a dynamical model having the property that a given
initial Hamiltonian evolves randomly---but isospectrally---in such a
way that the associated density function on the space of isospectral
Hamiltonians approaches a steady state distribution given by the
\textit{canonical ensemble}. Furthermore, we apply the resulting
equilibrium state to calculate thermal expectation values of other
observables, leading to new physical predictions in some limiting
(quenched and annealed) cases.

\begin{figure}[h]
\includegraphics[width=20pc]{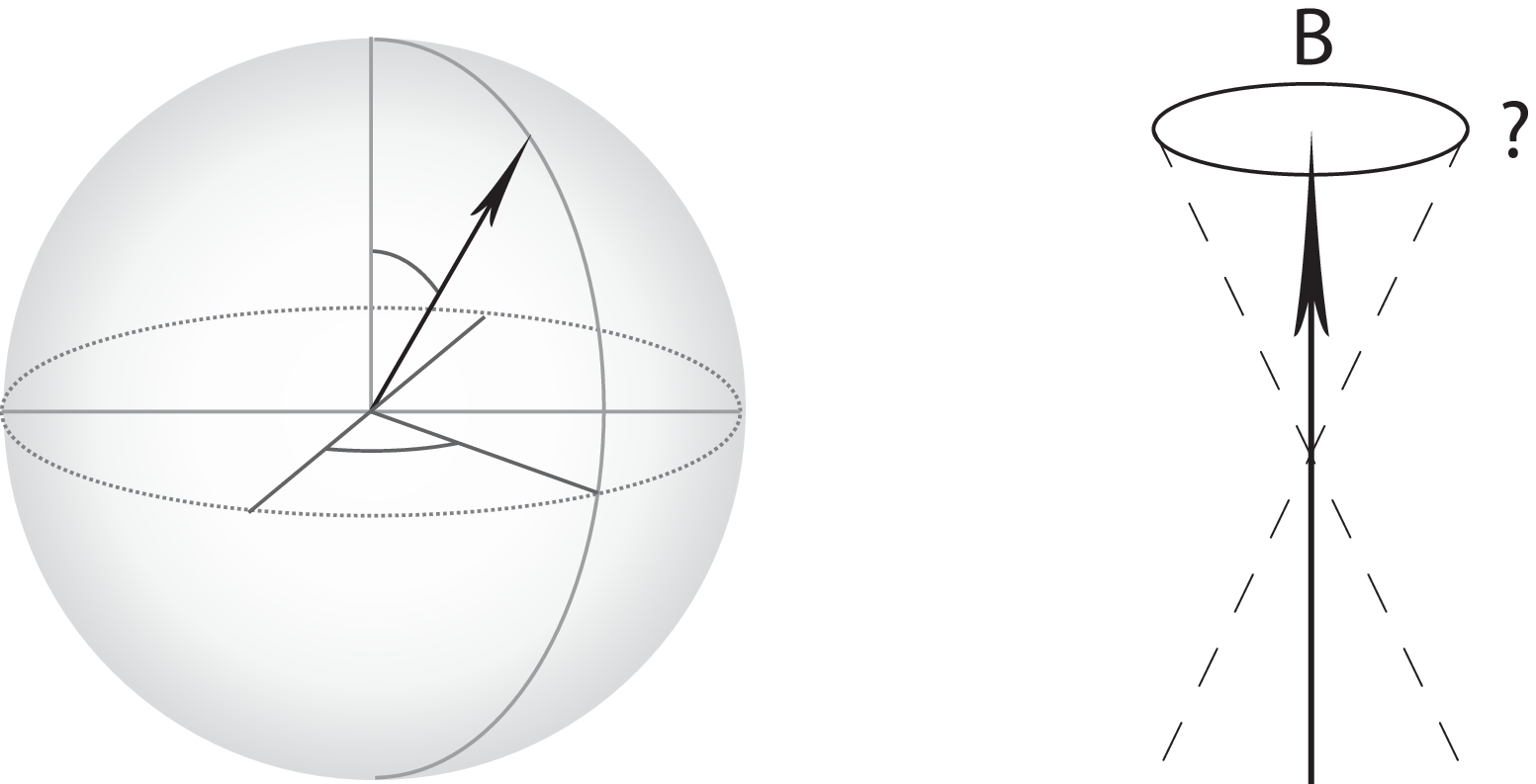}\hspace{1.2pc}%
\begin{minipage}[b]{16.5pc}
\caption{\label{fig:1}\textit{Spin in fluctuating magnetic field}.
The space of pure states of a spin-$\frac{1}{2}$ particle, in
external magnetic field ${\bf B}$, is the surface of the Bloch
sphere. In statistical theory of quantum mechanics the state is
represented by statistical distributions of the pure states, whereas
the magnetic field ${\bf B}$ that specifies the Hamiltonian is held
fixed. What happens if the direction of the field ${\bf B}$ is
itself subject to a small fluctuation? }
\end{minipage}
\end{figure}

\section{Approach to equilibrium}
\label{sec:2}

In classical statistical mechanics the notion of a \textit{gradient
flow} plays an important role in describing the approach to
equilibrium: A system immersed in a heat bath naturally tends to
release its energy into the environment and thus approach its
minimum energy state, and this tendency is characterised by a
Hamiltonian gradient flow. An equilibrium state is attained when
this flow is on average counterbalanced by thermal noise due to a
random interaction with the bath. Here the magnitude of the noise is
determined by the temperature of the bath. Accordingly, the idea we
are going to introduce here is a gradient flow equation on the space
of Hamiltonians with the property that the eigenstates of an
arbitrary initial Hamiltonian $H_0$ at time $t=0$ tend toward
alignment with those of a reference Hamiltonian, denoted by $G$.
Thus, $G$ plays the role of the `fixed' Hamiltonian in conventional
quantum statistical mechanics. The eigenstates of $H_t$ thus evolve
toward those of $G$ under the flow. By introducing a suitable noise
term, we are able to characterise the approach to an equilibrium
distribution.

The dynamical model for characterising approach to equilibrium is
given by
\begin{eqnarray}
\frac{\rd H_t}{\rd t} = -\lambda \left[H_t,[H_t,G]\right] + [H_t,
\omega_t], \label{eq:1}
\end{eqnarray}
where $\lambda\in{\mathds R}_+$. Here $\{\omega_t\}$ denotes a
skew-symmetric matrix of independent white noise terms, and
$[H_t,\omega_t]$ is the Lie bracket of these with $H_t$ (see also
\cite{brockett2}). Hence $[H_t,\omega_t]$ is symmetric and linear in
both $H_t$ and $\omega_t$. The term $\left[H_t,[H_t,G]\right]$ gives
rise to the aforementioned gradient flow in the space of
Hamiltonians. The Hermitian matrix $G$ plays the role of the
`Hamiltonian of the Hamiltonians' in the sense that $G$ determines
the motion in the space of Hamiltonians. In particular, we can
regard the linear function $\tr(HG)$ on the space of the totality of
Hermitian matrices $H$ as representing the `energy' function defined
on that space.

There is an invariant measure (stationary solution) associated with
the evolutionary equation (\ref{eq:1}). This is given by the
canonical density:
\begin{eqnarray}
\rho(H) \propto \exp\big(-\lambda\,{\rm tr}(HG)\big), \label{eq:2}
\end{eqnarray}
which can be used as a new basis for studying quantum statistical
mechanics. In units $\hbar=1$ the coupling $\lambda$ has dimension
$[{\rm Energy}]^{-1}$, and has the interpretation of representing
the inverse temperature for the `Hamiltonian bath' (not to be
confused with the thermal bath in which the system, and possibly
also the apparatus determining the Hamiltonian, is immersed). The
canonical density (\ref{eq:2}) can also be derived by entropy
maximisation subject to the constraint that the energy $\tr(HG)$ has
a definite expectation value.

\section{The gradient flow: double bracket equation}
\label{sec:3}

Let us begin by examining properties of the gradient term in
(\ref{eq:1}). Specifically, consider the following dynamical
equation for Hermitian matrices:
\begin{eqnarray}
\frac{\rd H_t}{\rd t} = -\lambda \left[H_t,[H_t,G]\right] .
\label{eq:3}
\end{eqnarray}
Note that in terms of the Hermitian matrix $X=\ri [H,G]$ the
double-bracket evolution (\ref{eq:3}) can be rewritten as
\begin{eqnarray}
\frac{\rd H}{\rd t} = \ri\lambda[H,X], \label{eq:4}
\end{eqnarray}
which formally is just the Heisenberg equation of motion. However,
owing to  the $H$-dependence of $X$ the evolution is nonunitary. The
Hamiltonians $H_0$ and $G$ are both assumed nondegenerate. The flow
induced by (\ref{eq:3}) satisfies the following properties: (i) the
evolution is isospectral, i.e. the eigenvalues of $H_0$ are
preserved; and (ii) the evolution gives the `alignment'
$\lim_{t\to\infty} [H_t,G]=0$.

We remark that the double bracket flow was first introduced in the
context of magnetism (Landau-Lifshitz equation) \cite{landau}. In
its modern form it was introduced by Brockett \cite{brockett} and
has been successfully applied to many areas, such as optimal
control, linear programming, sorting algorithms, and dissipative
systems (see references cited in \cite{BEH}).

In the case of a $2\times2$ Hamiltonian the gradient flow equation
(\ref{eq:3}) can be solved straightforwardly. In this case we
express the Hamiltonian in terms of the Pauli matrices:
\begin{eqnarray}
H_t=\half\,u_t{\mathds 1} +\half\,\nu\, {\boldsymbol{\sigma}}
\!\cdot\! {\mathbf n}_t, \label{eq:7}
\end{eqnarray}
where ${\mathbf n}_t =({\rm x}_t,{\rm y}_t,{\rm z}_t)$. Similarly
for the reference Hamiltonian $G$ we write
\begin{eqnarray}
G=\half\,v{\mathds 1} +\half\,\mu\, {\boldsymbol{\sigma}}
\!\cdot\!{\mathbf g} \label{eq:8}
\end{eqnarray}
for a unit vector ${\mathbf g}$. A calculation shows that the
solution to (\ref{eq:3}) reads
\begin{eqnarray}
H_t = \half\! \left(\! \begin{array}{cc} u_0\!-\!\nu \tanh(\omega
t-c_0) & \nu\,{\rm sech}(\omega t-c_0)\,\re^{-{\rm i} \phi_0} \\
\nu\,{\rm sech}(\omega t-c_0)\,\re^{{\rm i} \phi_0} & u_0 \!+\! \nu
\tanh( \omega t-c_0) \end{array} \!\right)\!,
\end{eqnarray}
where $\omega=\lambda\nu\mu$, and $c_0=\tanh^{-1}(\cos\theta_0)$ and
$\phi_0$ are initial values  \cite{BEH}. Furthermore, the
eigenvalues of $H_t$ are time-independent, and we have
\begin{eqnarray}
\lim_{t\to\infty} H_t = \half \left( \begin{array}{cc} u_0-\nu & 0
\\ 0 & u_0+\nu \end{array} \right). \label{eq:55}
\end{eqnarray}
Thus, the Hamiltonian is asymptotically diagonalised in the
$G$-basis. Observe that ${\rm tr}H_t$ and ${\det}H_t$ are conserved
quantities. Therefore, the flow induced by (\ref{eq:3}) for fixed
initial values $u_0$ and $|{\bf n}_0|$ is confined to a two-sphere
${\mathcal L}$, which is isomorphic to the state space of a
two-level system. It follows that in two dimensions we have the
equivalence of the Schr\"odinger and Heisenberg pictures, even
though the dynamical equation is not unitary. Since $u_0$ and $|{\bf
n}_0|$ are constant, we fix these and focus our attention on
${\mathcal L}$ parameterised by the dynamical coordinates
$(\theta_t,\phi_t)$.

\begin{figure}[h]
\includegraphics[width=11pc]{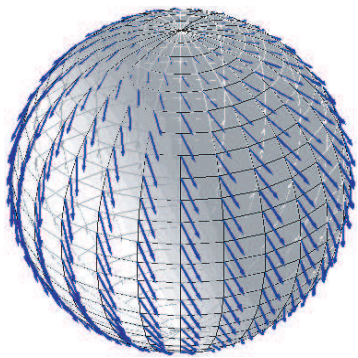}\hspace{4pc}%
\begin{minipage}[b]{23pc}
\caption{\textit{Gradient flow with unitary motion on the sphere
${\mathcal L}$}. The
  vector field generated by the unitarily-modified gradient-flow
  (\ref{eq:88}) is plotted. The first term in (\ref{eq:88}) generates
  a rotation around the $G$-axis (which is chosen to be the $z$ axis
  here), while the second term generates geodesic flows toward the
  south pole.
  The axis ${\mathbf n}_0$ of the initial Hamiltonian $H_0$ spirals
  around the $G$-axis ${\mathbf g}$ and is asymptotically aligned with
  the latter (south pole in this example).
  \label{fig:2}}
\end{minipage}
\end{figure}

We remark that the dynamical equation (\ref{eq:3}) can be modified
to include a unitary term:
\begin{eqnarray}
\frac{\rd H_t}{\rd t} = -\ri[H_t,G] - \lambda \left[H_t,[H_t,G]
\right] \label{eq:88}
\end{eqnarray}
without greatly affecting its physical characteristics. In the
$2\times2$ example, the only change occurs in the phase so that
instead of $\phi_t=\phi_0$ we have $\phi_t =\phi_0+\mu t$.

\section{Elements of stochastic differential geometry}
\label{sec:4}

We now wish to introduce a Brownian term into the deterministic flow
(\ref{eq:3}). Specifically, we consider a uniform Brownian field on
the isospectral subspace of the space of Hermitian matrices (the
sphere ${\mathcal L}$ in the $2\times2$ case). Before we proceed,
however, it will be useful to recall how stochastic motions can be
defined on a manifold. The basic process we consider is the Wiener
process $\{W_{t}\}$ defined on a filtered probability space
$({\Omega}, {\cal F},{\mathbb P})$. Here ${\Omega}$ is the sample
space, ${\cal F}$ is a $\sigma$-field on ${\Omega}$, and ${\mathbb
P}$ is the probability measure. The filtration of ${\mathcal F}$
determines the causal structure of $({\Omega}, {\cal F},{\mathbb
P})$. This is given by a parameterised family $\{{\mathcal
F}_t\}_{0\leq t<\infty}$ of nested $\sigma$-subfields satisfying
${\mathcal F}_s\subset {\mathcal F}_t\subset {\mathcal F}$ for any
$s\leq t<\infty$. We say that $\{W_{t}\}$ is a Wiener process if:
(a) $W_{0}=0$; and (b) $\{W_t\}$ is Gaussian such that
$W_{t+h}-W_{t}$ has mean zero and variance $|h|$ (see \cite{hida}).
A process $\{\sigma_{t}\}$ is said to be adapted to the filtration
$\{{\cal F}_{t}\}$ generated by $\{W_{t}\}$ if its random value at
time $t$ is determined by the history of $\{W_{t}\}$ up to that
time.

If $\{\sigma_{t}\}$ is ${\cal F}_{t}$-adapted, then the stochastic
integral $M_{t} = \int_{0}^{t} \sigma_{s} \rd W_{s}$ exists,
provided that $\{\sigma_{t}\}$ is almost surely square-integrable.
If the variance of $\{M_{t}\}$ exists, then $\{M_{t}\}$ satisfies
the \textit{martingale} conditions ${\mathbb E}[|M_{t}|]<\infty$ and
${\mathbb E}[M_{t}|{\cal F}_{s}]=M_{s}$, where ${\mathbb E}[-]$
denotes expectation with respect to the measure ${\mathbb P}$. The
latter condition implies that given the history of the Wiener
process up to time $s$ the expectation of $M_t$ for $t\geq s$ is
given by its value at $s$.

A general \textit{Ito process} is defined by an integral of the form
\begin{eqnarray}
x_{t}=x_{0}+\int_{0}^{t}\mu_{s}\rd s+\int_{0}^{t}\sigma_{s}\rd
W_{s}, \label{eq:20}
\end{eqnarray}
where $\{\mu_{t}\}$ and $\{\sigma_{t}\}$ are called the drift and
the volatility of $\{x_{t}\}$. A convenient way of expressing
(\ref{eq:20}) is to write $\rd x_{t}=\mu_{t}\rd t+\sigma_{t}\rd
W_{t}$, and to regard the initial condition $x_0$ as implicit. In
the special case $\mu_{t}=\mu(x_t)$ and $\sigma_t= \sigma(x_{t})$,
where $\mu(x)$ and $\sigma(x)$ are prescribed functions, the process
$x_t$ is said to be a diffusion.

This analysis can be generalised to the case of a diffusion
$\{x_{t}\}$ taking values on a manifold ${\mathfrak M}$, driven by
an $m$-dimensional Wiener process $\{W_{t}^{i}\}_{i=1,\ldots,m}$.
Let $\nabla_{a}$ be a torsion-free connection on ${\mathfrak M}$
such that for any vector field $\xi^a$ its covariant derivative in
local coordinates is
\begin{eqnarray}
\delta_{\bf b}^b \delta_a^{\bf a}(\nabla_b\xi^a) = \frac{\partial
\xi^{\bf a}}{\partial x^{\bf b}} + \Gamma_{\bf bc}^{\bf a}\xi^{\bf
c},
\end{eqnarray}
where $\delta_a^{\bf a}$ is the standard coordinate basis in a given
coordinate patch. Suppose we have an Ito process taking values in
${\mathfrak M}$. Let $x_t^{\bf a}$ denote the coordinates of the
process in a particular patch. Then writing $h^{\bf ab}
=\sigma_i^{\bf a}\sigma^{{\bf b}i}$ we define the drift process
$\mu^{\bf a}$ by
\begin{eqnarray}
\mu^{\bf a}\rd t= \rd x^{\bf a} +\half \Gamma_{\bf bc}^{\bf a}
h^{\bf bc} \rd t - \sigma_i^{\bf a}\rd W_t^i. \label{eq:72}
\end{eqnarray}
Alternatively, we can write the \textit{covariant Ito differential}
as $\rd x^a = \delta_{\bf a}^a ( \rd x^{\bf a}+\half \Gamma_{\bf
bc}^{\bf a} h^{\bf bc} \rd t)$, where $\delta^a_{\bf a}$ is the dual
coordinate basis. Then (\ref{eq:72}) can be represented as
\begin{eqnarray}
\rd x^{a}= \mu^{a} \rd t+\sigma^{a}_{i}\rd W^{i}_{t}.
\end{eqnarray}
If $\mu^{a}(x)$ and $\sigma^{a}_{i}(x)$ are $m+1$ vector fields on
${\mathfrak M}$, then the general diffusion process on ${\mathfrak
M}$ is governed by a stochastic differential equation $\rd x^{a}=
\mu^{a}(x) \rd t+\sigma^{a}_{i}(x)\rd W^{i}_{t}$, where $\rd x^{a}$
is the covariant Ito differential associated with the given
connection (see Hughston \cite{hughston}).

For the characterisation of the diffusion process it suffices to
specify a connection on ${\mathfrak M}$, and a metric is not
required. The quadratic relation $\rd x^{a}\rd x^{b}=h^{ab}\rd t$,
where $h^{ab}=\sigma_{i}^{a} \sigma^{bi}$, follows from the Ito
identities $\rd t^2=0$, $\rd t\rd W_t^i=0$, and $\rd W_t^i \rd
W_{t}^{j}= \delta^{ij}\rd t$. Then for any smooth function $\phi(x)$
on ${\mathfrak M}$ we define the associated process
$\phi_{t}=\phi(x_{t})$, and Ito's formula takes the form
\begin{eqnarray}
\rd\phi_t&=&(\nabla_a\phi)\rd x^a +\half(\nabla_a\nabla_b\phi) \rd
x^a \rd x^b \nonumber \\ &=& \left(\mu^{a}\nabla_{a}\phi+\half
h^{ab} \nabla_{a}\nabla_{b}\phi\right)\rd t +
\sigma^{a}_{i}\nabla_{a}\phi \rd W^i_t . \label{eq:26}
\end{eqnarray}
The probability law for $x_t$ is characterised by a density function
$\rho(x,t)$ on ${\mathfrak M}$ that satisfies the Fokker-Planck
equation
\begin{eqnarray}
\frac{\partial\rho}{\partial t}=-\nabla_{a}(\mu^{a}\rho)+ \half
\nabla_{a}\nabla_{b}(h^{ab}\rho). \label{eq:4.7}
\end{eqnarray}
The diffusion is said to be nondegenerate if $h^{ab}$ is of maximal
rank. If $g_{ab}$ is a Riemannian metric on ${\mathfrak M}$ and
$\nabla_{a}$ is the associated Levi-Civita connection, then if
$h^{ab}=\sigma^{2}g^{ab}$, the process $x_t$ is a Brownian motion
with drift on ${\mathfrak M}$, with volatility parameter $\sigma$.

\section{Diffusion model for thermalisation}
\label{sec:5}

Consider a stochastic differential equation of the form
\begin{eqnarray}
\rd x^a = \mu^a \rd t + \kappa \sigma^a_i \rd W_t^i \label{eq:sde}
\end{eqnarray}
on a real manifold ${\mathfrak M}$. Here $\kappa$ is a constant, the
drift $\mu^a$ is a vector field on ${\mathfrak M}$, and the vectors
$\{\sigma^{a}_{i}\}$ constitute an orthonormal basis in the tangent
space of ${\mathfrak M}$. In this case the associated Fokker-Planck
equation reads
\begin{eqnarray}
\frac{\partial}{\partial t}\,\rho_t(x) = -\nabla_{a}(\mu^{a}\rho_t)
+ \half \kappa^{2}\nabla^{2}\rho_t . \label{eq:29}
\end{eqnarray}
For our model we require that the drift vector $\mu^a$ represent the
double-bracket gradient flow (\ref{eq:3}). This is achieved by
choosing
\begin{eqnarray}
\mu^a=-\half\,\kappa^2 \lambda \nabla^a G,
\end{eqnarray}
where $G(x)$ is a function on ${\mathfrak M}$ given by $\tr(HG)$.
Then it follows that there exists a unique stationary solution to
(\ref{eq:29}), given by the canonical density
\begin{eqnarray}
\rho(x) = \frac{\exp(-\lambda G(x))}{\int_{\mathfrak M}\exp(-\lambda
G(x))\rd V}. \label{eq:31}
\end{eqnarray}
If ${\mathfrak M}$ is the space of pure states, then we have a model
for thermalisation of quantum states introduced by Brody \& Hughston
\cite{brody2}.

\section{Two-dimensional case in more detail}
\label{sec:6}

To illustrate these results in more explicit terms we consider a
system consisting of a single spin-$\frac{1}{2}$ particle immersed
in an external magnetic field. The Hamiltonian is then $H=-{\mathbf
B} \cdot {\mathbf S}$, where ${\mathbf B}$ denotes the field and
${\mathbf S}$  the spin vector. The direction of the field ${\mathbf
B}$, however, is subject to fluctuations around its stable
direction, specified by $G$ (directed along the $z$-axis). For the
dynamical equation we obtain \cite{BEH}:
\begin{eqnarray}
\left\{ \begin{array}{l} \rm d\theta_t = \omega\, \sin\theta_t \rd t
+ \sqrt{2\nu}(\rd W_t^1+\rd W_t^2) \\ \rd \phi_t =
-\frac{1}{\sin\theta_t}\sqrt{2\nu} (\rd W_t^1-\rd W_t^2).
\end{array} \right. \label{eq:33}
\end{eqnarray}
The associated Fokker-Planck equation reads
\begin{eqnarray}
{\dot\rho} = -\omega(\cos\theta + \sin\theta\,
\partial_\theta) \rho + 2 \nu{\textstyle{\left(
\partial_\theta^2\!+ \frac{1}{\sin^2\theta}\,
\partial_\phi^2 \right)}}\rho, \label{eq:34}
\end{eqnarray}
where $\partial_\theta=\partial/\partial\theta$ and $\partial_\phi
=\partial/\partial\phi$. The asymptotic solution is the canonical
density function:
\begin{eqnarray}
\rho(\theta,\phi)= \frac{\lambda\mu}{2\pi\sinh(\frac{1}{2}
\lambda\mu)}\,\exp\left(-\half\lambda\mu\cos\theta\right).
\label{eq:35}
\end{eqnarray}
Direct substitution shows that (\ref{eq:35}) is the stationary
solution to (\ref{eq:34}).

\begin{figure}[h]
\includegraphics[width=20pc,height=10.0pc]{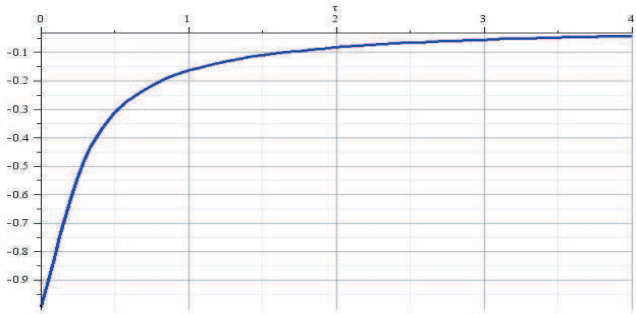}\hspace{2pc}%
\begin{minipage}[b]{16pc}
\caption{\label{fig:3}The plot of the expectation $\langle
\cos\theta\rangle_\lambda$ as a function of the inverse Hamiltonian
temperature $\tau=1/\lambda$. For $\tau\ll1$ we have $\langle
\cos\theta\rangle_\lambda\simeq-1$, whereas for $\tau\gg1$ we find
$\langle \cos\theta\rangle_\lambda\simeq0$.}
\end{minipage}
\end{figure}

It follows from (\ref{eq:35}) and the use of the volume element $\rd
V=\frac{1}{4} \sin\theta \rd\theta\rd\phi$ that the equilibrium mean
Hamiltonian is
\begin{eqnarray}
\langle H\rangle = \half \left( \begin{array}{cc} u_0+\nu\langle
\cos\theta\rangle_\lambda & 0 \\ 0 & u_0-\nu\langle
\cos\theta\rangle_\lambda, \end{array} \right), \label{eq:36}
\end{eqnarray}
where
\begin{eqnarray}
\langle \cos\theta\rangle_\lambda = \frac{2}{\lambda\mu}- \frac{1}
{\tanh(\frac{1}{2} \lambda\mu)}.
\end{eqnarray}
We may regard the parameter $\lambda$ as representing the `inverse
temperature' for the Hamiltonian. If the noise level is high
($\lambda\ll1$), then the direction of the external field ${\mathbf
B}$ on the average lies close to the $xy$-plane so that $\langle
\cos\theta\rangle_\lambda \simeq0$. If the noise level is low
$\lambda\gg1$, then the field ${\mathbf B}$ on the average is
parallel to the $z$-axis and we have $\langle
\cos\theta\rangle_\lambda\simeq-1$. We plot $\langle
\cos\theta\rangle_\lambda$ as a function of $\tau=1/\lambda$ (see
Fig.~\ref{fig:3}).

\section{Quantum statistical mechanics of disordered systems}
\label{sec:7}

Now we consider how the statistical theory of Hamiltonians presented
above can be applied to quantum statistical mechanics, when the
system and the apparatus specifying the Hamiltonian are both
immersed in a heat bath with inverse temperature $\beta$. In this
context it is natural to borrow ideas from the spin glass
literature. We may take the averaged Hamiltonian $\langle
H\rangle_\lambda$ as the starting point of  the analysis---this
gives the analogue of an \textit{annealed} average:
\begin{eqnarray}
\langle O\rangle_A = \frac{{\rm tr} \left(O\re^{-\beta \langle
H\rangle_\lambda}\right)}{{\rm tr} \left(\re^{-\beta \langle
H\rangle_\lambda}\right)}
\end{eqnarray}
of an observable $O$. Such an averaging, however, will change the
eigenvalues of $H$.

Alternatively, we may use the `unaveraged' Hamiltonian to compute
the expectation of an observable $O$, and then take its average:
\begin{eqnarray}
\langle O\rangle_Q = \left\langle \frac{{\rm tr} (O\re^{-\beta
H})}{{\rm tr} (\re^{-\beta H})}\right\rangle_\lambda.
\end{eqnarray}
This gives the analogue of a \textit{quenched} average. The
canonical quenched average of the Hamiltonian $G=\sigma_z$ is
\begin{eqnarray}
\langle G\rangle_Q = \half \mu\, {\textstyle \tanh\left(
\half\beta\nu\right) \left( \frac{1}{\tanh\left( \frac{1}{2}
\lambda\mu\right)}- \frac{2}{\lambda\mu} \right)},
\end{eqnarray}
whereas the canonical annealed average of $G$ is
\begin{eqnarray}
\langle G\rangle_A = \half \mu\, {\textstyle \tanh\left[
\half\beta\nu \left( \frac{1}{\tanh\left( \frac{1}{2}
\lambda\mu\right)}-\frac{2}{\lambda\mu} \right) \right]}.
\end{eqnarray}
These results suggest a new line of studies on the extended quantum
statistical mechanics of disordered systems. The plot below shows
the annealed (blue) and quenched (red) averages of $\sigma_z$, as a
function of the bath temperature $T=1/\beta$, for fixed $\lambda$
such that $\lambda^{-1}=0.1$.
\begin{figure}[h]
\includegraphics[width=18pc,height=10.0pc]{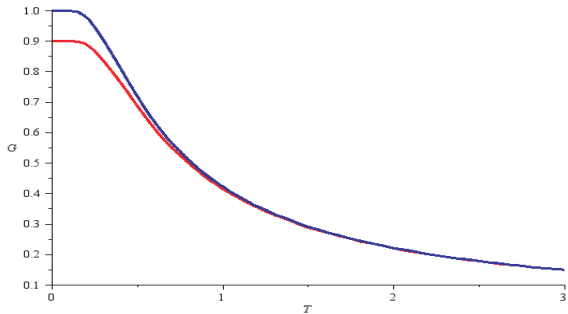}\hspace{2pc}%
\begin{minipage}[b]{16pc}
\caption{\label{fig:4}\textit{Quenched and annealed averages of
$G$}. The
  functions $\langle G\rangle_Q$ and $\langle G\rangle_A$ are
  plotted against the temperature $T=\beta^{-1}$, where we set
  $\lambda^{-1}=0.1$, $v=0$, $\nu=1$, and $\mu=2$ so that
  $G=\sigma_z$.
  The `quenched magnetisation' $\langle\sigma_z\rangle_Q$ does not
  attain the maximum value $1.0$ at zero temperature unless
  $\lambda^{-1}=0$.}
\end{minipage}
\end{figure}

\section{Examination of higher-dimensional cases}
\label{sec:8}

The geometry of higher-dimensional Hermitian matrices is somewhat
more intricate than the two-dimensional case examined above. The
space of $N\times N$ Hermitian matrices has the structure of the
product ${\mathds R}^N \times \mathbb{CP}^N \times \mathbb{CP}^{N-1}
\times \cdots \times \mathbb{CP}^{1}$, where $\mathbb{CP}^{k}$
denotes the complex projective $k$-space. This space is considerably
larger than the space $\mathbb{CP}^N$ of pure states upon which
$N\times N$ Hermitian matrices act, and as a consequence the
equivalence of the Schr\"odinger and Heisenberg pictures for a
nonunitary motion would in general be lost.

A $3\times3$ Hermitian matrix $H$ can be written as
$H=E_1|E_1\rangle\langle E_1|+E_2|E_2\rangle\langle E_2|
+E_3|E_3\rangle\langle E_3|$, where $\{E_i\}$ are the eigenvalues
and $\{|E_i\rangle\}$ are the associated eigenstates. The degrees of
freedom for the energy eigenvalues correspond to the open space
${\mathds R}^3$; the remaining degrees of freedom corresponding to
$\mathbb{CP}^2 \times \mathbb{CP}^{1}$ are encoded in the
specification of the energy eigenstates. Writing $\{|g_i\rangle\}$
for the eigenstates of $G$, we can express $|E_1\rangle$ in the
form:
\begin{eqnarray}
|E_1\rangle = \sin\half\vartheta\cos\half\varphi |g_1\rangle +
\sin\half\vartheta\sin\half\varphi\re^{{\rm i}\xi} |g_2\rangle +
\cos\half\vartheta\re^{{\rm i}\eta} |g_3\rangle,
\end{eqnarray}
which determines a point in $\mathbb{CP}^{2}$. There is a
$\mathbb{CP}^{1}$ worth degrees of freedom left for $|E_2\rangle$:
\begin{eqnarray}
|E_2\rangle &=&
\left(\cos\half\alpha\sin\half\vartheta\cos\half\varphi-\sin\half
\alpha \sin\half\varphi\re^{{\rm i}\beta}\right)|g_1\rangle
\nonumber \\ && + \left(\cos\half\alpha\cos\half\vartheta \sin
\half\varphi\re^{{\rm i}\xi} +\sin\half \alpha
\cos\half\varphi\re^{{\rm i}(\beta+\xi)} \right)|g_2\rangle +
\cos\half\alpha \sin\half\vartheta\re^{{\rm i}\eta} |g_3\rangle.
\end{eqnarray}
The specifications of $|E_1\rangle$ and $|E_2\rangle$ leave no
further freedom left for $|E_3\rangle$ and we have
\begin{eqnarray}
|E_3\rangle &=&
\left(\sin\half\alpha\cos\half\vartheta\cos\half\varphi+\cos\half
\alpha \sin\half\varphi\re^{{\rm i}\beta}\right)|g_1\rangle
\nonumber \\ && + \left(\sin\half\alpha\cos\half\vartheta \sin
\half\varphi\re^{{\rm i}\xi} -\cos\half\alpha \cos\half\varphi
\re^{{\rm i}(\beta+\xi)} \right)|g_2\rangle + \sin\half\alpha
\sin\half\vartheta\re^{{\rm i}\eta} |g_3\rangle.
\end{eqnarray}
In this manner we obtain the nine parameters required for the
specification of an arbitrary $3\times3$ Hermitian matrix.

It should be remarked that the foregoing procedure is merely an
example of how one might parameterise a generic $N\times N$
Hermitian matrix in a systematic manner; the scheme is somewhat
unconventional in that a generic $2\times2$ Hermitian matrix in this
parameterisation is written
\begin{eqnarray}
H = \omega_+ \left( \begin{array}{cc} 1 & 0 \\ 0 & 1 \end{array}
\right)+\omega_- \left(\begin{array}{cc} \cos\theta & \sin\theta\,
\re^{-{\rm i}\phi} \\ \sin\theta\, \re^{{\rm i}\phi} & -\cos\theta
\end{array} \right),
\end{eqnarray}
where $\omega_\pm = \frac{1}{2}(E_1\pm E_2)$. This set of
coordinates is nevertheless convenient because it isolates invariant
quantities $\omega_\pm$ from the coordinates $(\theta,\phi)$ of the
isospectral submanifolds.

For our application in quantum statistical mechanics of disordered
systems we are required to determine the partition function
\begin{eqnarray}
Z(\lambda) = \int \re^{-\lambda\tr(HG)} \rd V. \label{eq:33}
\end{eqnarray}
In the case of a $3\times3$ Hamiltonian, since the dynamical
equation (\ref{eq:1}) preserves the three eigenvalues of $H$, the
dynamical motion stays on the isospectral submanifold spanned by the
six angular variables $\vartheta,\varphi,\alpha\in[0,\pi]$ and
$\xi,\eta,\beta\in[0,2\pi]$, with volume element
\begin{eqnarray}
\rd V = {\textstyle\frac{1}{128}} \sin\alpha \sin\vartheta
(1-\cos\vartheta) \sin\varphi \, \rd\alpha \, \rd\beta \,
\rd\vartheta\, \rd \varphi\, \rd \xi\, \rd \eta.
\end{eqnarray}
Since we are working in the $G$-basis, the calculation of $\tr(HG)$
is straightforward, and the integral (\ref{eq:33}) reduces to an
expression analogous to the integral representation for a Bessel
function.

The choice of parameterisation adapted here for a generic $N\times
N$ Hermitian matrix need not be the most adequate for our purpose.
An alternative approach is to write $H=U^\dagger E U$, where $E$ is
a diagonal matrix with eigenvalues $\{E_i\}$, and $U$ is a unitary
matrix. Then the calculation of the partition function becomes
semi-Gaussian, and we are left with an integration over the
invariant Haar measure \cite{sugiura}. Ideas from matrix theory or
random matrices might prove useful in the statistical analysis of
disordered quantum systems introduced here.

\ack The authors thank R.~Brockett and E.~J.~Brody for comments and
stimulating discussions. DDH thanks the Royal Society of London for
partial support by its Wolfson Merit Award.


\vskip10pt

\begin{thebibliography}{99}


\bibitem{brockett2} Brockett,~R.~W. ``Notes on stochastic
processes on manifolds'' In {\em Systems and Control in the
Twenty-First Century} (C.~Byrnes, \textit{et al}., eds.) pp. 75-101,
(Boston: Birkh\"{a}user, 1997).

\bibitem{landau} Landau,~L.~D. and Lifshitz,~E.~M.
``On the theory of the dispersion of magnetic permeability in
ferromagnetic bodies'' {\em Phys. Z. Sowietunion} \textbf{8} 153-169
(1935).

\bibitem{brockett} Brockett,~R.~W. ``Dynamical systems that sort
lists, diagonalise matrices, and solve linear programming problems''
{\em Lin. Alg. Appl.} \textbf{146} 79-91 (1991).

\bibitem{BEH} Brody,~D.~C., Ellis, D.C.P., and Holm, D.D.
``Hamiltonian statistical mechanics'' {\em J. Phys.} A\textbf{41}
502002 (2008).

\bibitem{hida} Hida,~T. {\em Brownian Motion}. (Berlin, Germany:
Springer, 1980).

\bibitem{hughston} Hughston,~L.~P. ``Geometry of stochastic state
vector reduction'' {\em Proc. Roy. Soc. London} A\textbf{452}
953-979 (1996).

\bibitem{brody2} Brody,~D.~C. and Hughston,~L.~P. ``Thermalisation
of quantum states'' {\em J. Math. Phys.} \textbf{40} 12-18 (1999).

\bibitem{sugiura} Sugiura,~M. {\em Unitary Representations and
Hamonic Analysis} (Amsterdam: North-Holland, 1990).

\end{thebibliography}
\end{document}